# A Framework for Auditing Multilevel Models using Explainability Methods

Debarati Bhaumik[1], Diptish Dey[1], Subhradeep Kayal[2]
[1]Amsterdam University of Applied Sciences, Amsterdam, The Netherlands
[2]Erasmus University Medical Center, Rotterdam, The Netherlands
d.bhaumik@hva.nl
d.dey2@hva.nl
s.kayal@erasmusmc.nl

**Abstract**: Multilevel models (MLMs) are increasingly deployed in industry across different functions. Applications usually result in binary classification within groups or hierarchies based on a set of input features. For transparent and ethical applications of such models, sound audit frameworks need to be developed. In this paper, an audit framework for technical assessment of regression MLMs is proposed. The focus is on three aspects: model, discrimination, and transparency & explainability. These aspects are subsequently divided into sub-aspects. Contributors, such as inter MLM-group fairness, feature contribution order, and aggregated feature contribution, are identified for each of these sub-aspects. To measure the performance of the contributors, the framework proposes a shortlist of KPIs, among others, intergroup individual fairness ($Diff_{Ind\_MLM}$) across MLM-groups, probability unexplained *(PUX)* and percentage of incorrect feature signs *(POIFS)*. A traffic light risk assessment method is furthermore coupled to these KPIs. For assessing transparency & explainability, different explainability methods (SHAP and LIME) are used, which are compared with a model intrinsic method using quantitative methods and machine learning modelling.

Using an open-source dataset, a model is trained and tested and the KPIs are computed. It is demonstrated that popular explainability methods, such as SHAP and LIME, underperform in accuracy when interpreting these models. They fail to predict the order of feature importance, the magnitudes, and occasionally even the nature of the feature contribution (negative versus positive contribution on the outcome). For other contributors, such as group fairness and their associated KPIs, similar analysis and calculations have been performed with the aim of adding profundity to the proposed audit framework. The framework is expected to assist regulatory bodies in performing conformity assessments of AI systems using multilevel binomial classification models at businesses. It will also benefit providers, users, and assessment bodies, as defined in the European Commission's proposed Regulation on Artificial Intelligence, when deploying AI-systems such as MLMs, to be future-proof and aligned with the regulation.

**Keywords:** auditable AI, multilevel model, explainability, discrimination, ethics

## 1. Introduction

Classification models in machine learning (ML) using logistic regression are common, albeit the higher accuracy reported in some literature of its well-studied counterpart random forest (Begenilmis and Uskudarli, 2018; Maroco et al, 2011). Conjointly, logistic regression and MLMs (Gelman and Hill, 2006) could be greatly beneficial in circumstances, where a model needs to demonstrate local behaviour while recognizing the global context. Applications result in binary classification within groups or hierarchies inherent in the dataset. MLMs are deployed across different industries such as insurance (Amu et al, 2022), food production (Rodrigues, 2019), and entertainment (Perano et al, 2021) within functions such as marketing and supply chain management (Carter et al, 2015; Awaysheh et al, 2021).

The unfavourable impact of ML models on society is common knowledge (Angwin et al, 2016). Mikians et al (2012) presented their findings on price discrimination by online retailers based upon geographical location. More recently, algorithmic pricing in hospitality and tourism and its social impact has been researched by van der Rest et al (2022). Decisions made by classification models are interpretable to the extent to which an observer can understand the rationale behind those decisions (Biran and Cotton, 2017) and is able to consequently provide explanations. The need for explainability is driven by among others, examination and scientific explanation (Miller, 2019). The importance of a more structured and actionable approach towards explainability is significant. (Raji and Buolamwini, 2019; Kazim, Denny and Koshiyama, 2021). The need to monitor social responsibility and ethical behaviour (Dunlop, 1998; Verschoor, 1998) in the context of corporate governance, as opposed to, competitiveness (Ho, 2005) in the market is unprecedented.

An industry in which the use of statistical/ML models is highly regulated is Banking, where supervision through legislation (Bank, 2019) supported by stringent audit practices is common practice today. The European Central Bank (Bank, 2019) provides an overview of practices required for the use of internal credit risk models. It is noteworthy that ML models deployed in Banking are based upon tabular data and therefore could be less complex than applications elsewhere. Accordingly, this paper proposes an audit framework for use in logistic regression MLMs.

## 2. Elements of audit

What is tantamount to fairness, bias and/or discrimination, is a subject of discussion in contemporary literature. The broad range of stakeholders, including individuals, corporations, legislators, psychologists, and social activists (Schroepfer, 2021; Kordzadeh and Ghasemaghaei, 2022), justifies the challenge of auditing ML algorithms (Raji and Buolamwini, 2019; Landers and Behrend, 2022). Xue, Yurochkin and Sun (2020) presented a set of inferential tools from an auditor's perspective, whereas Landers and Behrend (2022) present a framework for evaluating bias. They discuss the challenges faced by auditors and present the components of an Artificial Intelligence (AI) system to be audited as model-related, information & perception-related, and meta-components. Simultaneously, these issues are legislatively reflected in a structured and actionable manner within the proposed AI Act (Commissie, 2021), which presents a conformity regime. Offering a tiered risk-based approach, it addresses "remote biometric identification" and "biometric categorisation". Furthermore, it sets out an enforcement mechanism through audits and conformity assessments while recognizing the significance of competence in specific AI technologies. As opposed to the EU data protection act, GDPR, (Voigt and Von dem Bussche, 2017), which caters to compliance needs of data controllers when processing personal data, the proposed EU AI Act addresses accountability in the whole supply chain and focusses also on models that process this data.

### 2.1 Significance of the proposed EU AI act

The proposed EU AI Act encompasses a number of aspects, among others, classification of 'high-risk AI systems,' prohibited practices, transparency obligations, governance, and compliance procedures. It "*sets out the legal requirements for high-risk AI systems in relation to data and data governance, documentation and recording keeping, transparency and provision of information to users, human oversight, robustness, accuracy and security*" (Commissie, 2021, p. 13). Articles 10 to 15 of Chapter 2 Title III of the proposed act elaborates on these legal requirements.

Combining these legal requirements of the proposed act with components extracted from academic literature, at an aggregated level, we propose 5 audit aspects: model, discrimination, transparency & explainability, IS/IT systems, and processes. These aspects and their consequent sub-aspects, contributors and KPIs are presented in Table 1 and explained in subsequent sections. *IS/IT systems* and *processes* are beyond the scope of this paper. The sub-aspects listed in Table 1 are non-exhaustive and are selected based on their significance in the proposed AI act, their relevance to MLMs, and choices made to keep the discussion focussed.

### 2.2 Audit elements in focus

Auditing models entail concentrating assessment efforts on assumptions and considerations made prior to and amidst building the model, and performance levels post deployment. The range of sub-aspects audited would include design choices, data collection methods, *formulation of relevant assumptions*, availability & suitability of data sets (pre-training), level of accuracy, robustness, and sensitivity & stability. A selection of these sub-aspects and their corresponding contributors, as prescribed in Table 1, are evaluated in the context of MLMs in section 3. When auditing *discrimination* aspects, we focus our efforts in section 4 on the examination of possible biases through the sub-aspects: *group & individual fairness*. In section 5, we address *transparency* and *explainability* by assessing and demonstrating the accuracy of commonly known explainable methods. Consequently, sections 3, 4 and 5 present KPIs that substantiate the audit framework, making it actionable.

Additionally, the audit framework is demonstrated through a case-study using an open-source US health insurance dataset (Kaggle, 2018). The dataset consists of 1338 rows with the columns: age, gender, BMI, number of children, smoker, region, and insurance charges. A binary classification problem is formulated for predicting which insureds are entitled to an insurance claim of greater than $6,000 per region based on the independent variables age, BMI, and number of children.

## 3. Model aspect

Auditing *model* aspect includes assessment of various sub-aspects that are related to model-building, such as model assumptions, statistical properties, accuracy of predictions, model robustness, and others. This section presents an assessment of two sub-aspects: *formulation of relevant assumptions*, and *accuracy of predictions*.

Table 1: Audit framework (* indicates beyond the scope of this paper, ** indicates proposed values)

| Aspects | Sub-aspects | Contributors | KPIs | Traffic light-RAG score | | |
|---|---|---|---|---|---|---|
| | | | | Red | Amber | Green |
| Model | Formulation of relevant assumptions | Model assumptions | Extent and depth of their presence in technical documentation | Qualitative | Qualitative | Qualitative |
| | | Statistical properties | VIF | > 5.0 | [1.0-5.0] | < 1.0 |
| | | | SWT | $p < 0.05$ | $0.05<p<0.1$ | $p > 0.1$ |
| | | | BPT | $p < 0.05$ | $0.05<p<0.1$ | $p > 0.1$ |
| | Accuracy of predictions | Discriminatory power | AUC-ROC | [0.0-0.5) | [0.5-0.8) | [0.8-1.0] |
| | | Predictive power | F1-score | [0.0-0.5) | [0.5-0.8) | [0.8-1.0] |
| Discrimination | Group fairness | Predicted outcome | SP | $\gg 0.2$ | $\gtrapprox 0.2$ | < 0.2 |
| | | | DI | $\ll 0.8$ | $\lessapprox 0.8$ | (0.8-1.0] |
| | | Predicted & actual outcome | EqualOdds** | > 0.2 | [0.1-0.2] | < 0.1 |
| | Individual fairness | Intra MLM-group fairness | $Diff_{Ind}$ | $\gg 0.2$ | $\gtrapprox 0.2$ | < 0.2 |
| | | Inter MLM-group fairness | $Diff_{Ind\_MLM}$ | $\gg 0.2$ | $\gtrapprox 0.2$ | < 0.2 |
| Transparency & explainability | Accuracy of explainability methods | Feature contribution order | $\rho_{order}$ | [-1.0- 0.3) | [0.3-0.8) | [0.8-1.0] |
| | | Aggregated feature contribution | PUX** | > 0.2 | [0.1 – 0.2] | < 0.1 |
| | | Feature contribution sign | POIFS** | [100-20)% | [20-10)% | [10-0]% |
| | Sensitivity* | Sensitivity of accuracy KPIs to background dataset* | | | | |
| | | Sensitivity of model output around inflection points* | | | | |
| Processes* | | | | | | |
| IS/IT systems* | | | | | | |

Contributors to the *formulation of relevant assumptions* include *model assumptions* and *statistical properties*. Model assumptions differ per AI/statistical model (Hastie et al, 2009). Furthermore, most ML algorithms assume that training, validation, and test datasets follow the same statistical distribution. KPIs for auditing *model assumptions* are qualitative measures of the extent and depth of their presence in the technical documentation. Various *model assumptions* are examined by testing *statistical properties* of a model and its data, e.g., to check the independence of independent variables, a test for multicollinearity is performed. Most *statistical properties* are model-dependent, e.g., to check if the residuals in generalized linear models are normally distributed and have constant variance, homoskedasticity is tested. This is not a requirement for tree-based models.

For classification problems, contributors to the *accuracy of predictions* include *discriminatory power*, and *predictive power* of a model. *Discriminatory power* of a model is the ability to differentiate between the groups being assessed and *predictive power* is the measure of its goodness of fit.

## 3.1 Model assumptions of logistic regression MLMs

In logistic regression MLMs used for binary classification (Gelman and Hill, 2006), model parameters (coefficients) can vary per MLM-group. Random variables are used to model the variation between the MLM-groups. The assumption that the random effects come from a common distribution (resulting in MLM-groups sharing information), improves the predictive power of MLMs in comparison with simple regression (linear or logistic) models. The random effects are modelled within the regression coefficients: in the slopes as fixed effects, in the intercepts as random effects, or in the slopes & intercepts as mixed effects.

A simple logistic regression MLM with three independent variables $x^1$, $x^2$, $x^3$ and the dependent variable $y$ can be represented as:

$$\log\left(\frac{\mathbb{P}(y_i = 1)}{1 - \mathbb{P}(y_i = 1)}\right) = \alpha_{[j]i} + \beta^1_{[j]i} x^1_i + \beta^2_{[j]i} x^2_i + \beta^3_i x^3_i + \epsilon_i \quad (1)$$

where, $i = \{1, \ldots, N\}$, $N$ is the total number of data points, $j = \{1, \ldots, J\}$, $J$ is the total number of MLM-groups, $\mathbb{P}(y_i = 1)$ is the probability that the $i^{th}$ dependent variable $y_i$ belongs to class 1, $\alpha_{[j]i}$ is the intercept term varying per MLM-group j, $\beta^1_{[j]i}$ and $\beta^2_{[j]i}$ are the slope term varying per MLM-group $j$, $\beta^3_i$ is the slope coefficient which remains constant per MLM-group, $x^1_i, x^2_i$, and $x^3_i$ are the $i^{th}$ independent variable and $\epsilon_i$ is the error term. The $\alpha$'s and $\beta's$ are the model coefficients to be estimated. Assumptions of the model in (1) are:

- The log-odds, $\log\left(\frac{\mathbb{P}(y_i = 1)}{1 - \mathbb{P}(y_i = 1)}\right)$, has a linear relationship with the independent variables $x_i$.
- The independent variables $(x)$ are mutually uncorrelated.
- The error terms are normally distributed with mean zero.
- The varying intercept and slope terms follow normal distributions.
- The training and testing datasets follow the same distribution.

These assumptions lead to the following *statistical properties*:

- $\epsilon_i \sim \mathcal{N}(0, \sigma_y^2)$,
- $\alpha_{[j]} \sim \mathcal{N}(\mu_\alpha, \sigma_\alpha^2)$, and $\beta_{[j]} \sim \mathcal{N}(\mu_\beta, \sigma_\beta^2)$, for $j = 1, \ldots, J$.

To the case study, a MLM as per equation (1) is fitted with 4 MLM-groups corresponding to the variable region (northeast, northwest, southeast, and southwest). The intercept ($\alpha$), and the slopes corresponding to age ($\beta^1$) and BMI ($\beta^2$) per MLM- group are varied, whereas the slope corresponding to the number of children ($\beta^3$) constant. The dataset had a train-test split of 95%-5%. The model was trained to estimate the model parameters.

## 3.2 KPIs for assessing statistical properties

For assessing *statistical properties* three KPIs are recommended: (i) *variance inflation factor (VIF)* (Hastie et al, 2009), and (ii) *Shapiro-Wilk test (SWT)* (Shapiro and Wilk, 1965), and (iii) *Breusch-Pagan test (BPT)* (Cook and Weisberg, 1983).

VIF measures the degree of collinearity in data by estimating how much of the variance of a coefficient in a linear model increases due to shared variance with other independent variables compared to the case where the variables are uncorrelated. A value of VIF > 5.0 generally indicates high collinearity among variables that needs to be mitigated (Schwarz, Chapman and Feit, 2020). The SWT, which is used to check normality in given samples, are applied on the errors or residuals ($\epsilon$ term in equation (1)) in MLMs. If these residuals do not have constant variance (homoskedasticity), then the outcomes of MLM cannot be trusted. The BPT checks for heteroscedasticity (lack of constant variance) in the residuals.

The computed KPI values for the case-study are as follows:

- *VIF*: The VIF for age, BMI, and number of children are estimated to be 7.5, 7.8 and 1.8 respectively. These imply that collinearity of age and BMI need to be mitigated.
- *SWT*: The test of residuals has a *p-value* $\approx 0.0$. This implies that there exists some non-linear relationship between the independent and the dependent variables.
- *BPT*: The test for the residuals has a *p-value* $\approx 0.0$. This implies that the residuals do not have constant variance.

### 3.3 KPIs for assessing accuracy of predictions

Area under the ROC curve (*AUC-ROC*) and *F1-score* are recommended as KPIs for measuring the discriminatory power and the predictive power of a MLM (Tharwat, 2020). The *F1-score*, commonly used to measure performance of a classifier, has a range [0,1], with 1 indicating a perfect classifier. Area under the ROC curve, with a range [0,1], measures separability of two classes in a binary classifier. A high value of ~1 indicates a good classifier; 0.5 implies random classifications; < 0.5 implies very poor performance. (Tharwat, 2020). In the case-study, the KPIs are computed as follows:

- Average *F1-score* is 0.83 for all MLM-groups, implying that the binary classifier possesses a good predictive power.
- Average *AUC-ROC* score is 0.91 for all MLM-groups. This is greater than 0.85, implying that the binary classifier possesses a good discriminatory power.

## 4. Auditing discrimination

In a backdrop of expanding benefits of AI applications, the prevailing notion that algorithms taking decisions are objective, is to a large extent driven by the subjectivity of humans. The lack of past legislative urgency in assessing the ethical consequences of such decisions is due to a common belief that these algorithms are far from human-like subjectivity. Underscoring the importance of researching machine ethics, Anderson and Anderson (2007) discuss that work in ethical theory is often too isolated from actual applications in which ethical dilemmas arise. One such dilemma is bias, which needs to be researched in AI systems in the context of different applications.

Bias, synonymously used with discrimination and fairness in literature, in predictive analytics could be a result of several causes (Chouldechova and Roth, 2020). It could arise at different stages in the model development process: during pre-processing in the training dataset, during classification by the algorithm and at the post-processing stage (Haas, 2019). Due to their inherent trait, classification algorithms, e.g., in MLMs, learn patterns and make predictions based upon the available training data. A biased training dataset will certainly result in a biased prediction. Controlling the quality of the training dataset for biases and demonstrating the lack of bias through statistical comparisons with overlaying population is therefore essential.

### 4.1 Bias KPIs

Whereas measuring bias resulting from the pre-processing phase through a single KPI is challenging, KPIs for the subsequent stages are abundantly researched in literature. Such KPIs are either associated with group fairness or with individual fairness (Pessach and Shmueli, 2022; Haas, 2019) and can be further labelled as KPIs based upon predicted outcome, predicted & actual outcome, or predicted probabilities & actual outcome. The scope of subsequent discussions is limited to individual fairness and to the concepts of predicted outcome and predicted & actual outcome in group fairness.

#### 4.1.1 Group fairness – predicted outcome

The most commonly occurring KPIs in literature are *Statistical Parity (SP)* and *Disparate Impact (DI)*, which measure positive prediction across different groups. Such groups are characterized by attributes, such as gender and ethnicity. *SP* is calculated as a difference,

$$\left|\mathbb{P}[\hat{Y}=1|S=1] - \mathbb{P}[\hat{Y}=1|S \neq 1]\right| \leq \varepsilon \qquad (2)$$

whereas *DI* is calculated as a ratio:

$$\frac{\mathbb{P}[\hat{Y}=1|S=1]}{\mathbb{P}[\hat{Y}=1|S \neq 1]} \geq 1 - \varepsilon \qquad (3)$$

In the equations above, $\hat{Y} = 1$ corresponds to desired classification, such as getting an insurance or a job, $S$ represents the protected feature, such as gender or ethnicity, and $S = 1$ and $S \neq 1$ represent the privileged and under-privileged groups, respectively. For *DI* it is desirable to have the ratio as high as possible, since this would imply that a desired classification is similar across the groups. Similarly, for *SP*, the value should be as low as possible. Benchmarks for these KPIs are seldom, except for *DI* in the context of employment law in America (Stephanopoulos, 2018), where a value of 0.8 or higher is acceptable.

### 4.1.2 Group fairness - predicted & actual outcome

The KPIs based only on predicted outcome fall short of considering the actual outcomes that an otherwise accurate classification algorithm could have and thereby label the latter as unfair. The concept of *Equalized-Odds* (Hardt, Price and Srebro, 2016) overcomes this disadvantage by computing the differences between the false-positive rates and the true-positive rates (Pessach and Schmueli, 2022). Ideally, an algorithm is fair if the false-positive and the true-positive rates are equal in the privileged and under-privileged groups. The KPIs are calculated as follows:

$$Diff_{FPR} = \left| \mathbb{P}[\hat{Y} = 1 | S = 1, Y = 0] - \mathbb{P}[\hat{Y} = 1 | S \neq 1, Y = 0] \right| \leq \varepsilon \tag{4}$$

$$Diff_{TPR} = \left| \mathbb{P}[\hat{Y} = 1 | S = 1, Y = 1] - \mathbb{P}[\hat{Y} = 1 | S \neq 1, Y = 1] \right| \leq \varepsilon \tag{5}$$

The *Equalized-Odds* KPI is subsequently calculated as follows:

$$EqualOdds = 0.5 * (Diff_{FPR} + Diff_{TPR}) \tag{6}$$

In an ideal world, *Equalized-Odds* is zero. When applying the above KPIs to MLM-groups, it is useful to calculate the KPIs for each MLM-group and analyse them in their respective MLM-groups.

### 4.1.3 Individual fairness

As opposed to equal treatment of groups, individual fairness relates to outcomes for individuals. The underlying principle is that similar individuals are treated equally. In the context of MLMs, this equality can be interpreted within a MLM-group, as,

$$Diff_{Ind} = \left| \mathbb{P}(\hat{Y}^{(i)} = y | X^{(i)}, S^{(i)}) - \mathbb{P}(\hat{Y}^{(j)} = y | X^{(j)}, S^{(j)}) \right| \leq \varepsilon, \; if \; d(i,j) \approx 0 \tag{7}$$

where *i* and *j* denote two individuals with *S* and *X* representing sensitive attributes and associated features of the individuals respectively (Pessach and Schmueli, 2022). The distance metric $d(i,j) \approx 0$ ensures that the two individuals compared are in some respect similar. Equation (7) is extended to the interpretation of equality across various MLM-groups, as,

$$Diff_{Ind\_MLM} = \left| \mathbb{P}(\hat{Y}^{(i,a)} = y | X^{(i,a)}, S^{(i,a)}) - \mathbb{P}(\hat{Y}^{(j,b)} = y | X^{(j,b)}, S^{(j,b)}) \right| \leq \varepsilon \tag{8}$$

where *a* and *b* refer to two different MLM-groups. It is important to calculate and interpret $Diff_{Ind\_MLM}$, since MLM-groups could differ substantially in their regression models as discussed earlier in section 3.

## 4.2 Fairness calculations in the use-case

Group fairness: With gender (male/female) as the sensitive group, the following KPIs are computed.
- *SP* of 0.083, implying a relatively small difference between the sensitive groups.
- *DI* of 0.89, indicating a value within the acceptable bandwidth of 0.8 or higher.
- *Equalized-odds* per MLM-group of 0.28, 0.5, 0.33, 0.29 for northeast, northwest, southeast, and southwest, respectively. These values are too high and unacceptable.

Individual fairness:
- $Diff_{Ind}$ was computed as 0.18 for two very similar individuals in the MLM-group northwest, with age as the sensitive attributes and feature values (age = 30, BMI = 30, children = 2, gender = female) and (age = 35, BMI = 30, children = 2, gender = female). This is not a high difference in probability for a small difference in age.
- $Diff_{Ind\_MLM}$ was computed as 0.07 for two individuals with same characteristics (age = 30, BMI = 30 children = 1, gender = female) in two MLM-groups, northeast and southeast. The small value indicates a minor difference in fairness.

## 5. Transparency and explainability aspect

Transparent and explainable AI models help create trust and understand model outputs at scale by enabling *interpretability* (Lipton, 2018; Chen et al, 2022). The two main types of explanation methods are feature attribution explanations, which explains the importance of each feature to model predictions, and counterfactual explanations, which explains what minimum modification should be applied to the input data to get a desired output (Chen et al. 2022; Molnar, 2020). In this paper, the focus is on feature attribution explanations and the proposed sub-aspect: *accuracy of explanation methods*. Feature attribution explanations elucidate to what extent each feature has contributed to the model prediction; thereby, explaining the importance of each feature in that particular instance prediction (Zhou et al. 2021). The two most commonly used additive feature attribution methods are SHAP (SHapley Additive exPlanations) and LIME (local linear regression model) (Gramegna and Giudici, 2021; Ward et al, 2021).

The proposed contributors, *feature contribution order, aggregated feature contribution*, and *feature contribution sign,* are assessed by comparing SHAP and LIME with a model intrinsic method. To assess feature contribution order, *Spearman's rank correlation coefficient* ($\rho_{order}$) is used as a KPI (Hollander, Wolfe and Chicken, 1973). To assess aggregated feature contribution and feature contribution sign, the KPIs *probability unexplained* ($PUX$) and *percentage of incorrect feature signs* ($POIFS$) respectively are proposed.

### 5.1 Model explainability methods

For regression MLMs, model intrinsic feature attribution explanations are enabled through interpretability of the estimated model coefficients in equation (1). Since $\log\left(\frac{\mathbb{P}(y_i = 1)}{1 - \mathbb{P}(y_i = 1)}\right)$, follows a linear relationship with the independent variables $x's$ (see equation (1)), the magnitude and sign of the $\beta's$ show the linear contribution of each independent variable to the dependent variable in the log-odds space. Similarly, the $\alpha_{[j]}$ represents the base value contribution to the log-odds term when all the independent variables are set to zero.

LIME is a model-agnostic method, which provides explanations by learning a surrogate interpretable model locally around the prediction (Ribeiro, Singh and Guestrin, 2016). The surrogate interpretable models used are generally linear regression models with regularization or decision trees (Molnar, 2020). In this paper Linear LIME is used, which uses a linear regression model to locally approximate Logistic Regression MLM predictions. This choice is made to facilitate comparison of the feature contribution/importance estimated by Linear LIME with the model intrinsic method.

Based on Shapley values from game theory, SHAP is a method designed to provide explanations for individual prediction instances by assigning each feature with an importance value. Lundberg and Lee (2017) propose model agnostic and model specific SHAP approximations. In the case study, the model agnostic Kernel SHAP method in log-odds is deployed. Kernel SHAP connects Shapley values and linear LIME to compute the importance of each feature.

Python libraries developed by Lundberg and Lee, (2017) and Ribeiro, Singh and Guestrin, (2016) are used for computing the SHAP and LIME feature contributions, respectively.

### 5.2 KPIs for assessing accuracy of explanation methods

$\rho_{order}$ is computed by comparing the ranks of feature contribution magnitude of SHAP and LIME with model intrinsic method. It can take values in the range $[-1,1]$, varying between a worst-case negative association (-1) to a best-case positive association (+1). $PUX$ is calculated as absolute value of the difference between the probability estimates obtained from the model intrinsic method and those obtained from the SHAP or LIME methods, i.e., $|\mathbb{P}(y_i = 1)_{ModelIntrinsic} - \mathbb{P}(y_i = 1)_{SHAP/LIME}|$. $POIFS$ is calculated as the ratio of the number of times SHAP or LIME estimate the model feature contribution sign incorrectly as compared to the model intrinsic method with the total number of features, i.e., $\frac{\#incorrect\ feature\ signs}{\#total\ features} \times 100$. The ideal values of $PUX$ and $POIFS$ are zero. The means of $\rho_{order}$, $PUX$, and $POIFS$ over 50 randomly sampled instances from the training dataset is computed and the process is repeated 10 times to estimate the KPIs for each MLM-group. Following are the estimated KPIs with their standard deviations for the four MLM-groups (northeast, northwest, southeast, southwest) for SHAP:

- $\rho_{order}^{SHAP}$: $(0.86 \pm 0.07), (0.80 \pm 0.05), (0.80 \pm 0.09)$, and $(0.82 \pm 0.06)$. Across all MLM-groups, SHAP feature contribution magnitude ranks are not fully correlated with model intrinsic, implying that SHAP occasionally produces incorrect feature contribution magnitudes.

- $PUX^{SHAP}$: $(0.1 \pm 0.01)$, $(0.09 \pm 0.005)$, $(0.07 \pm 0.003)$, and $(0.09 \pm 0.01)$. Across all MLM-groups, there is a probability difference between model intrinsic and SHAP of 0.1 of an insured being in class 1, which is moderately significant in this use-case.
- $POIFS^{SHAP}$: $(9.2 \pm 1.7)\%$, $(3.1 \pm 1.04)\%$, $(7.2 \pm 2.06)\%$, and $(2.6 \pm 1.36)\%$. Across all MLM-groups, SHAP feature contribution signs do not completely match with model intrinsic signs, implying that there are instances where SHAP estimates feature contribution signs incorrectly.

The above statistics may fail to detect outliers, which is important in an audit. In Figure 1, two case examples of incorrect explanations produced by SHAP are presented. In Figure 1-(a), it is observed that for an instance [age = 35, BMI = 40, children = 3], in the group northwest, SHAP reports magnitude, order, and sign of feature contributions incorrectly. Figure 1-(b) is another instance where SHAP reports magnitude and order incorrectly.

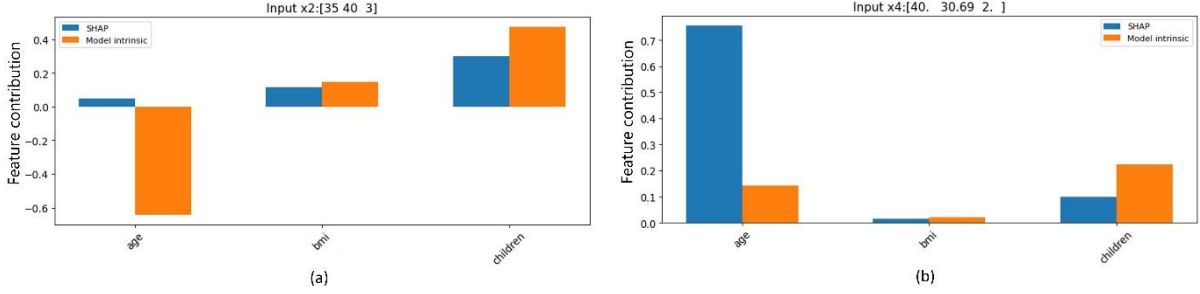

*Figure 1: Comparison of SHAP contributions with Model intrinsic in the MLM-group northwest*

Similarly, following are KPIs with their standard deviations for the four MLM-groups (northeast, northwest, southeast, and southwest) for LIME:
- $\rho_{order}^{LIME}$: $(0.92 \pm 0.01)$, $(0.83 \pm 0.04)$, $(0.72 \pm 0.03)$, and $(0.80 \pm 0.02)$.
- $PUX^{LIME}$: $(0.06 \pm 0.006)$, $(0.06 \pm 0.007)$, $(0.06 \pm 0.005)$, and $(0.05 \pm 0.008)$.
- $POIFS^{LIME}$: $(61 \pm 3.8)\%$, $(60 \pm 4.0)\%$, $(59 \pm 4.7)\%$, and $(60 \pm 3.7)\%$.

For $POIFS^{LIME}$, it is observed that LIME underperforms SHAP, and gets 60% of feature contribution signs wrong. However, $\rho_{order}^{LIME}$ is at par with SHAP, and LIME feature contribution magnitude ranks are not fully correlated with model intrinsic. $PUX^{LIME}$ performs better than $PUX^{SHAP}$ for this use-case. This may vary per use-case.

## 6. Discussion and conclusions

To audit logistic regression multilevel models, an audit framework with a set of assessment KPIs is proposed in Table 1. The framework focuses on the assessment of *model assumptions, accuracy of predictions, group & individual fairness*, and *accuracy of explainability methods*. The KPIs associated with each of these aspects are computed by fitting a model to an open-source dataset, and are assessed through a traffic light RAG-score.

The KPI for *model assumptions* is a qualitative measure of the extent and depth of the assumptions as reported in the technical documentation. The KPIs for assessing *statistical properties* are *variance inflation factor (VIF), Shapiro-Wilk test (SWT), and Breusch-Pagan test (BPT)*. The acceptable value for *VIF* is less than 1.0, for *SWT* and *BPT* is p > 0.1. The KPIs for assessing *accuracy of predictions* are *Area under the ROC curve* (*AUC-ROC*) and *F1-score*, which have both an acceptable range between [0.8, 1.0]. The KPIs for assessing *group fairness* are *Statistical Parity (SP), Disparate Impact (DI),* and *Equalized-Odds (EqualOdds),* with acceptable values of less than 0.2, between (0.8, 1.0], and less than 0.1, respectively. The KPIs for assessing *individual fairness* are *intra MLM-group fairness (Diff$_{Ind}$)* and *inter MLM-group fairness (Diff$_{Ind\_MLM}$),* both having an acceptable values of less than 0.2. The KPIs for assessing *accuracy of explainability methods* are *Spearman's rank correlation coefficient* ($\rho_{order}$), *probability unexplained* ($PUX$), and *percentage of incorrect feature signs* ($POIFS$), with acceptable values between [0.8-1.0], less than 0.1, and between [10-0]%, respectively.

For the case-study, a green RAG-score was obtained for the following KPIs: *AUC-ROC, F1-score, SP, DI, Diff$_{Ind}$, Diff$_{Ind\_MLM}$,* $PUX$ for both SHAP and LIME, and $POIFS$ for SHAP. An amber RAG-score was obtained for the following KPIs: $\rho_{order}$ for both SHAP and LIME. A red RAG-score was obtained for the following KPIs: *VIF, SWT, BPT,* and $POIFS$ for LIME. It is observed that even though the *statistical properties* of the model assumptions are not satisfied, the model scores high on the *accuracy of predictions.* From an explainability perspective, although it is observed that commonly used explainability methods occasionally produce incorrect feature contribution magnitudes and that SHAP performs better than LIME, SHAP fails to detect outliers (Figure 1). When stand-alone, these explainability methods fall short of providing trustworthy model output explanations. Also, the need for multiple KPIs to assess each aspect is evident, concluding that no single KPI or method is conclusive for any single aspect. For example, although the group fairness KPI, *Disparate Impact,* performs well, *Equalized-Odds* underperforms.

Future work includes testing and eventually extending the framework to other classification models. Aspects and sub-aspects, such as *processes, IS/IT systems*, and *sensitivity*, will also be explored. The framework presented in Table 1, when fully detailed, provides a structured and actionable approach to auditing AI systems.

## References

Amu, H., Dickson, K.S., Adde, K.S., Kissah-Korsah, K., Darteh, E.K.M. and Kumi-Kyereme, A., 2022. Prevalence and factors associated with health insurance coverage in urban sub-Saharan Africa: Multilevel analyses of demographic and health survey data. *Plos one*, *17*(3), p.e0264162.

Anderson, M. and Anderson, S.L., 2007. Machine ethics: Creating an ethical intelligent agent. *AI magazine*, *28*(4), pp.15-15.

Angwin, J., Larson, J., Mattu S., & Kirchner, L. (2016). Machine Bias. Retrieved April 24, 2022 from https://www.propublica.org/article/machine-bias-risk-assessments-in-criminal-sentencing

Bank, E.C., 2019. ECB guide to internal Models. Retrieved 01.02.2022: https://www.bankingsupervision.europa.eu/ecb/pub/pdf/ssm.guidetointernalmodels_consolidated_201910~97fd49fb08.en.pdf

Beğenilmiş, E. and Uskudarli, S., 2018, June. Organized behavior classification of tweet sets using supervised learning methods. In *Proceedings of the 8th International Conference on Web Intelligence, Mining and Semantics* (pp. 1-9).

Biran, O. and Cotton, C., 2017, August. Explanation and justification in machine learning: A survey. In *IJCAI-17 workshop on explainable AI (XAI)* (Vol. 8, No. 1, pp. 8-13).

Chen, V., Li, J., Kim, J.S., Plumb, G. and Talwalkar, A., 2022. Interpretable machine learning: Moving from mythos to diagnostics. *Queue*, *19*(6), pp.28-56.

Chouldechova, A. and Roth, A., 2020. A snapshot of the frontier of fairness in machine learning. *Communications of the ACM*, *63*(5), pp.82-89.

Cook, R.D. and Weisberg, S., 1983. Diagnostics for heteroscedasticity in regression. *Biometrika*, *70*(1), pp.1-10.

Commissie, E., 2021. Proposal for a Regulation of the European Parliament and of the Council laying down harmonised rules on Artificial Intelligence (Artificial Intelligence Act) and amending certain Union legislative acts. Retrieved 01.06.2022: https://eur-lex.europa.eu/legal-content/EN/TXT/?uri=CELEX%3A52021PC0206.

Dunlop, A. ed., 1998. *Corporate governance and control*. Kogan Page Publishers.

Gelman, A. and Hill, J., 2006. *Data analysis using regression and multilevel/hierarchical models*. Cambridge university press.

Gramegna, A. and Giudici, P., 2021. SHAP and LIME: an evaluation of discriminative power in credit risk. *Frontiers in Artificial Intelligence*, p.140.

Haas, C., 2019. The price of fairness - A framework to explore trade-offs in algorithmic fairness. In *40th International Conference on Information Systems, ICIS 2019*. Association for Information Systems.

Hardt, M., Price, E. and Srebro, N., 2016. Equality of opportunity in supervised learning. *Advances in neural information processing systems*, *29*.

Hastie, T., Tibshirani, R., Friedman, J.H. and Friedman, J.H., 2009. *The elements of statistical learning: data mining, inference, and prediction* (Vol. 2, pp. 1-758). New York: springer.

Ho, C.K., 2005. Corporate governance and corporate competitiveness: an international analysis. *Corporate Governance: An International Review*, *13*(2), pp.211-253.

Hollander, M., Wolfe, D.A. and Chicken, E., 2013. *Nonparametric statistical methods*. John Wiley & Sons.

Kaggle, 2018. *US Health Insurance Dataset*. [Online]. Retrieved 01.03.2022: https://www.kaggle.com/datasets/teertha/ushealthinsurancedataset

Kazim, E., Denny, D.M.T. and Koshiyama, A., 2021. AI auditing and impact assessment: according to the UK information commissioner's office. *AI and Ethics*, *1*(3), pp.301-310.

Kordzadeh, N. and Ghasemaghaei, M., 2022. Algorithmic bias: review, synthesis, and future research directions. *European Journal of Information Systems*, *31*(3), pp.388-409.

Landers, R.N. and Behrend, T.S., 2022. Auditing the AI auditors: A framework for evaluating fairness and bias in high stakes AI predictive models. *American Psychologist*.

Lipton, Z.C., 2018. The mythos of model interpretability: In machine learning, the concept of interpretability is both important and slippery. *Queue*, *16*(3), pp.31-57.

Lundberg, S.M. and Lee, S.I., 2017. A unified approach to interpreting model predictions. *Advances in neural information processing systems*, *30*.

Maroco, J., Silva, D., Rodrigues, A., Guerreiro, M., Santana, I. and de Mendonça, A., 2011. Data mining methods in the prediction of Dementia: A real-data comparison of the accuracy, sensitivity and specificity of linear discriminant analysis, logistic regression, neural networks, support vector machines, classification trees and random forests. *BMC research notes*, *4*(1), pp.1-14.

Mikians, J., Gyarmati, L., Erramilli, V. and Laoutaris, N., 2012, October. Detecting price and search discrimination on the internet. In *Proceedings of the 11th ACM workshop on hot topics in networks* (pp. 79-84).

Miller, T., 2019. Explanation in artificial intelligence: Insights from the social sciences. *Artificial intelligence*, *267*, pp.1-38.

Molnar, C., 2020. *Interpretable machine learning*. Lulu.com.

Perano, M., Casali, G.L., Liu, Y. and Abbate, T., 2021. Professional reviews as service: A mix method approach to assess the value of recommender systems in the entertainment industry. *Technological Forecasting and Social Change*, *169*, p.120800.

Pessach, D. and Shmueli, E., 2022. A Review on Fairness in Machine Learning. *ACM Computing Surveys (CSUR)*, *55*(3), pp.1-44.

Raji, I.D. and Buolamwini, J., 2019, January. Actionable auditing: Investigating the impact of publicly naming biased performance results of commercial ai products. In *Proceedings of the 2019 AAAI/ACM Conference on AI, Ethics, and Society* (pp. 429-435).

Ribeiro, M.T., Singh, S. and Guestrin, C., 2016, August. " Why should i trust you?" Explaining the predictions of any classifier. In *Proceedings of the 22nd ACM SIGKDD international conference on knowledge discovery and data mining* (pp. 1135-1144).

Rodrigues, D., Teixeira, R. and Shockley, J., 2019. Inspection agency monitoring of food safety in an emerging economy: A multilevel analysis of Brazil's beef production industry. *International Journal of Production Economics*, *214*, pp.1-16.

Schroepfer, M. (2021, March 11). *Teaching fairness to machines*. Facebook Technology. https://tech.fb.com/teaching-fairness-to-machines/

Schwarz, J.S., Chapman, C. and Feit, E.M., 2020. *Python for marketing research and analytics*. Springer Nature.

Shapiro, S.S. and Wilk, M.B., 1965. An analysis of variance test for normality (complete samples). *Biometrika*, *52*(3/4), pp.591-611.

Stephanopoulos, N.O., 2018. Disparate Impact, Unified Law. *Yale LJ*, *128*, p.1566.

Tharwat, A., 2020. Classification assessment methods. Applied Computing and Informatics, 17 (1), 168–192.

van der Rest, J.P., Sears, A.M., Kuokkanen, H. and Heidary, K., 2022. Algorithmic pricing in hospitality and tourism: call for research on ethics, consumer backlash and CSR. *Journal of Hospitality and Tourism Insights*, (ahead-of-print).

Verschoor, C.C., 1998. A study of the link between a corporation's financial performance and its commitment to ethics. *Journal of Business ethics*, *17*(13), pp.1509-1516.

Voigt, P. and Von dem Bussche, A., 2017. The eu general data protection regulation (gdpr). *A Practical Guide, 1st Ed., Cham: Springer International Publishing*, *10*(3152676), pp.10-5555.

Ward, I.R., Wang, L., Lu, J., Bennamoun, M., Dwivedi, G. and Sanfilippo, F.M., 2021. Explainable artificial intelligence for pharmacovigilance: What features are important when predicting adverse outcomes?. *Computer Methods and Programs in Biomedicine*, *212*, p.106415.

Xue, S., Yurochkin, M. and Sun, Y., 2020, June. Auditing ml models for individual bias and unfairness. In *International Conference on Artificial Intelligence and Statistics* (pp. 4552-4562). PMLR.

Zhou, J., Gandomi, A.H., Chen, F. and Holzinger, A., 2021. Evaluating the quality of machine learning explanations: A survey on methods and metrics. Electronics, 10(5), p.593.